\documentclass[11pt]{article}
\newcommand{\BE}{\begin{equation}}
\newcommand{\EE}{\end{equation}}
\newcommand{\BA}{\begin{eqnarray}}
\newcommand{\EA}{\end{eqnarray}}

\usepackage[dvips]{graphicx}
\usepackage{amssymb}

\begin{document}

\begin{titlepage}
\vspace*{24mm}
\begin{center}
   {\LARGE{\bf Comparison of Perturbative RG Theory \\ 
\vspace*{2mm}
with Lattice Data for the $4d$ Ising Model }}

\vspace*{22mm}

{\Large P. M. Stevenson}
\vspace*{3mm}\\
{\large T. W. Bonner Laboratory, Department of Physics and Astronomy \\
Rice University, P.O. Box 1892, Houston, TX 77251-1892, USA}
\vspace{25mm}\\
{\bf Abstract:}
\end{center}

Predictions for $(\phi^4)_4$ theory from renormalization-group-improved 
perturbation theory, as formulated by L\"uscher and Weisz, are compared to 
published (and some unpublished) data from lattice Monte-Carlo simulations 
of the $4$-dimensional Ising model.  Good agreement is found in all but one 
respect:---   the change in the wavefunction-renormalization constant 
$\hat{Z}_R$ across the phase transition is significantly greater than 
predicted.  A related observation is that propagator data in the broken 
phase show deviations from free-propagator form --- deviations that become 
larger, not smaller, closer to the continuum limit.  More data closer to 
the critical point are needed to clarify the situation.

\end{titlepage}

\newpage

\section{Introduction}

   A complete understanding of the $(\phi^4)_4$ theory is important not 
only as a fundamental problem in quantum field theory, but also for its 
implications for the Higgs mechanism.  According to conventional wisdom 
the continuum limit of lattice $(\phi^4)_4$ theory is described by a 
Renormalization Group (RG) analysis using RG functions calculated in 
perturbation theory.  The theory was developed in detail by Br\'ezin 
{\it et al} \cite{Brezin} and by L\"uscher and Weisz (LW) \cite{LW1,LW2}.
It predicts ``triviality'' in the sense that the renormalized coupling 
$g_R$ tends to zero.  

   In the late eighties LW's numerical predictions were compared with 
Monte-Carlo data for the 4-dimensional Ising model in both the symmetric 
phase \cite{MW, Montvay} and the broken phase \cite{Jansen}.  The aim of 
this paper is to revisit this comparison in the light of the much larger 
data set now available \cite{CCC,Balog,CCCS,CCunpub}.   

   One motivation for this exercise is the recent controversy between 
Balog, Duncan, Willey, Niedermayer and Weisz (BDWNW) \cite{Balog} and 
Cea, Consoli, and Cosmai (CCC) \cite{CCC}.  It is important to note that 
the raw data of the two groups agree very well; the dispute is solely over 
interpretation. I cannot pretend that my sympathies are neutral; for 
many years I have collaborated closely with CCC in gathering lattice 
Monte-Carlo evidence \cite{CCCS,Agodi} for an unconventional view of 
``triviality'' in $(\phi^4)_4$ theory advocated by Consoli and myself 
\cite{CS}.  Nevertheless, I intend here to take a detached view and, 
except for a few remarks in Sect. 5, I shall not discuss the ideas of 
Ref. \cite{CS}.  No change in my position is implied; I simply want to focus 
here on a limited question: {\it how well does perturbative RG theory agree 
with all the available lattice data?} 

    While BDWNW and CCC both describe fits of their data to formulas 
based on 1- or 2-loop perturbation theory, neither group has made a 
comprehensive comparison to the full LW theory incorporating 3-loop and 
higher-twist effects.  That exercise is performed here in Sect. 3, 
considering all measured quantities in the both the broken and symmetric 
phases.  Good agreement is found, except for the wavefunction-renormalization 
constant $\hat{Z}_R$, as discussed in Sect. 4.  Broken-phase propagator data 
are examined in Sect. 5.  BDWNW's data show the same deviation from 
free-propagator behaviour found in Ref. \cite{CCCS}.  Because of this 
deviation, different strategies for extracting ``mass'' and ``$\hat{Z}$'' 
parameters can lead to very different conclusions.  Non-Ising data are briefly 
considered in Sect. 6 and conclusions are summarized in Sect. 7.  
An appendix gives a detailed discussion of the data used.      

\section{Basic definitions}

    I shall use the notation of LW and BDWNW and I refer the reader to 
those papers for full definitions.  
Only a few key facts will be outlined here.   The lattice action for the 
$\phi^4$ theory is written as 
\BE
S= \sum_x \left[ - 2 \kappa \sum_{\mu=1}^{4} \phi(x) \phi(x+\hat{\mu}) + 
\phi(x)^2 + \lambda \left( \phi(x)^2-1 \right)^2 \right], 
\EE
which is equivalent to the more traditional expression
\BE
S= \sum_x \left[ \frac{1}{2} \sum_{\mu=1}^{4} 
\left( \partial_\mu \phi_0(x) \right)^2 + \frac{1}{2} m_0^2 \, \phi_0(x)^2 + 
\frac{g_0}{4!} \, \phi_0^4 \right],
\EE
where $\partial_\mu \phi_0(x) = \phi_0(x+\hat{\mu})-\phi_0(x)$.  The 
translation between the two formulations is given by 
\BE
\phi_0 = \sqrt{2 \kappa} \, \phi, \quad \quad 
m_0^2=\frac{(1-2 \lambda)}{\kappa}-8, \quad \quad 
g_0=\frac{6 \lambda}{\kappa^2}.
\EE
LW also define another parameter $\bar{\lambda}$ that varies between $0$ 
and $1$ as $\lambda$ ranges from $0$ to $\infty$.  The limit 
$\lambda \to \infty$ ($\bar{\lambda}=1$) corresponds to the Ising model, 
where $\phi(x)$ can take only the values $\pm 1$.    For a given 
$\bar{\lambda}$ there is a critical $\kappa$  
separating the symmetric and broken phases.  As $\kappa \to \kappa_c$ the 
correlation length (inverse of the physical mass in lattice units) 
diverges, according to the RG theory, so that $\kappa \to \kappa_c$ 
corresponds to the continuum limit.  

    The $\sqrt{2 \kappa}$ factor between the Ising field $\phi$ and the 
canonical field $\phi_0$ is the source of several notational nuisances.  
In particular, the field renormalization constant $Z_R$ defined by LW 
includes both the trivial $2 \kappa$ factor and the dynamical effects.  
I follow BDWNW in defining $\hat{Z}_R= 2 \kappa Z_R$ as the canonical 
field renormalization that obeys $\hat{Z}_R <1$.  The renormalized coupling 
constant $g_R$ and renormalized mass $m_R$ are defined as in LW, as are 
the susceptibility $\chi$ and vacuum expectation value 
$v=\langle \phi \rangle$.  To plot the data for the latter quantities I have 
first removed the power-law dependence on $(\kappa-\kappa_c)$ by defining
\BE
\tilde{\chi} \equiv \chi (\kappa-\kappa_c), \quad \quad 
\tilde{v}^2 = v^2/(\kappa-\kappa_c).
\EE
Notice that the combination $\chi v^2$ is ``dimensionless'' in this sense. 

    The RG theory involves coupled differential equations containing three 
RG functions $\beta$, $\gamma$ and $\delta$.  LW provide a recipe for 
constructing these functions to 3-loop order, including power-suppressed (or 
``higher twist'') scaling violations to 1-loop order.  The higher-twist 
terms, while negligible in the limit $\kappa \to \kappa_c$, are important 
further away from $\kappa_c$.  Because of them the integration of the 
differential equations must be done numerically.  Using {\it Mathematica}, 
I have implemented LW's procedure exactly as described in Refs. \cite{LW1}, 
\cite{LW2} (referred to below as LW(I) and LW(II), respectively).
\footnote{Copies of my {\it Mathematica} programs are available upon request.}
The LW procedure depends upon three integration constants $C_1, C_2, C_3$ 
as well as on the assumed value for $\kappa_c$. I have verified that my 
program precisely reproduces the results in the LW tables when the same input 
parameters are used.  
(The $C_i$ constants are defined in the symmetric phase; in the broken phase 
LW define corresponding constants $C_1'$, $C_2'$, $C_3'$ and then prove that 
$C_1'={\rm e}^{1/6} C_1$, while $C_2'=C_2$ and $C_3'=C_3$.  I shall quote 
only numerical values for the $C_i$'s, but of course the conversion from 
$C_1$ to $C_1'$ is taken into account in my program.)
  
    In the Ising case a quite precise value for $\kappa_c$ is known 
\cite{Stauffer}:
\BE
\kappa_c=0.074848(2)
\EE
and will be adopted here.  This value is consistent with earlier estimates, 
$0.074834(15)$ \cite{Gaunt} and $0.074851(8)$ \cite{Kenna}.  [In fact, I 
initially made fits using $0.074834$ and then tried $0.074851$ and found 
a small but distinct improvement.  On closer inspection, the improvement 
seemed to have slightly ``overshot'' and I was experimenting with slightly 
smaller ``compromise'' values when I became aware of the result of Ref. 
\cite{Stauffer}.  Thus, I adopt that value both because it has the smallest 
quoted uncertainty and because it seems to produce the best fits.]  

\section{Fits to lattice data}

    Table 1 of LW(II) gives predictions for the three integration constants 
based on LW(I)'s analysis that matched the RG procedure to the 
``high-temperature'' (small $\kappa$) expansion in the symmetric phase.  
For the Ising model these predictions are 
\BE
\label{LWparam}
\ln C_1=1.5(2), \quad\quad  \ln C_2=1.87(1), \quad\quad  \ln C_3=-3.0(1).
\EE
One can indeed fit the available symmetric-phase data quite well with 
parameters in this range (see later).  However, these parameter values do not 
yield a good fit to the broken-phase data.  This observation is in accord with 
the experience of CCC \cite{CCC}.  However, BDWNW \cite{Balog} point out 
that, with hindsight, the uncertainties quoted by LW may have been 
over-optimistically small, especially in the Ising case.  Also, LW used a 
much cruder approximate value for $\kappa_c$ ($0.07475(7)$) which is another 
source of the discrepancies found by CCC.  By adjusting the values of 
$\kappa_c$ and the $C_i$'s, BDWNW claim one can fit the broken-phase data 
very well with the conventional RG theory.  To check this assertion, I have 
considered all available Monte-Carlo Ising data (see Appendix) and made a 
number of fits, adjusting the parameters by trial and error.  Fig. 1 shows 
the best fit I have obtained, which uses the following parameter values:   
\BE
\label{fitparam}
\ln C_1=1.24, \quad\quad  \ln C_2=1.83, \quad\quad  \ln C_3=-2.90.
\EE
Indeed, this fit seems to be an entirely satisfactory description of the 
broken-phase data, given the theoretical uncertainties, especially at the 
larger $\kappa$'s.

      Comparing these parameter values with the LW values quoted above one 
sees a lower $\ln C_1$ value -- in accord with a remark in BDWNW 
that a value $\sim 1.2$ is needed.  However, note also the lower $\ln C_2$ 
value, which will be crucial in what follows.   

     [It is hard to quote meaningful uncertainties on the ``best fit'' 
parameter values given above.  The effects of varying the $C_i$'s are 
highly correlated, difficult to describe, and often hard to understand 
intuitively.  Also, one should give more weight to fitting the data points 
closer to $\kappa_c$, where the theoretical uncertainties are less.  
As a very rough guide I would say that changing any of the 
$\ln C_i$ values in Eq. (\ref{fitparam}) by plus or minus $0.06$, $0.01$, 
$0.01$, respectively, would lead to a discernible deterioration in the 
quality of the fit and changes by twice these amounts would be unacceptable.] 

    However, while the $C_i$ values in Eq. (\ref{fitparam}) yield an 
excellent fit to the broken-phase data, they do {\it not} yield a good fit 
to the symmetric-phase data; see Fig. 2.  The fits to $m_R$ and $g_R$ 
are quite good, but the theoretical curve for $\hat{Z}_R$ lies far below 
the data points.  That fact is a direct consequence of the lower $C_2$ value, 
since $\hat{Z}_R$ has a factor of $C_2$.    

    If one increases $\ln C_2$ to $1.862$ (back in accord with LW's 
predicted value) one can then obtain a good fit to symmetric phase data; 
see Fig. 3.  However, this change spoils the fit to the broken-phase 
data, not only for $\hat{Z}_R$, but also for $\tilde{\chi}$ and 
$\tilde{v}$; see Fig. 4.  I have tried adjusting the $C_i$'s further but I 
cannot find any ``compromise'' values that would make the problem go 
away.  One can fit either the broken-phase data, or the symmetric-phase data 
--- but one cannot fit both well simultaneously.

\linespread{1}


\begin{figure}[htp]
\begin{center}
\includegraphics[width=14.2cm]{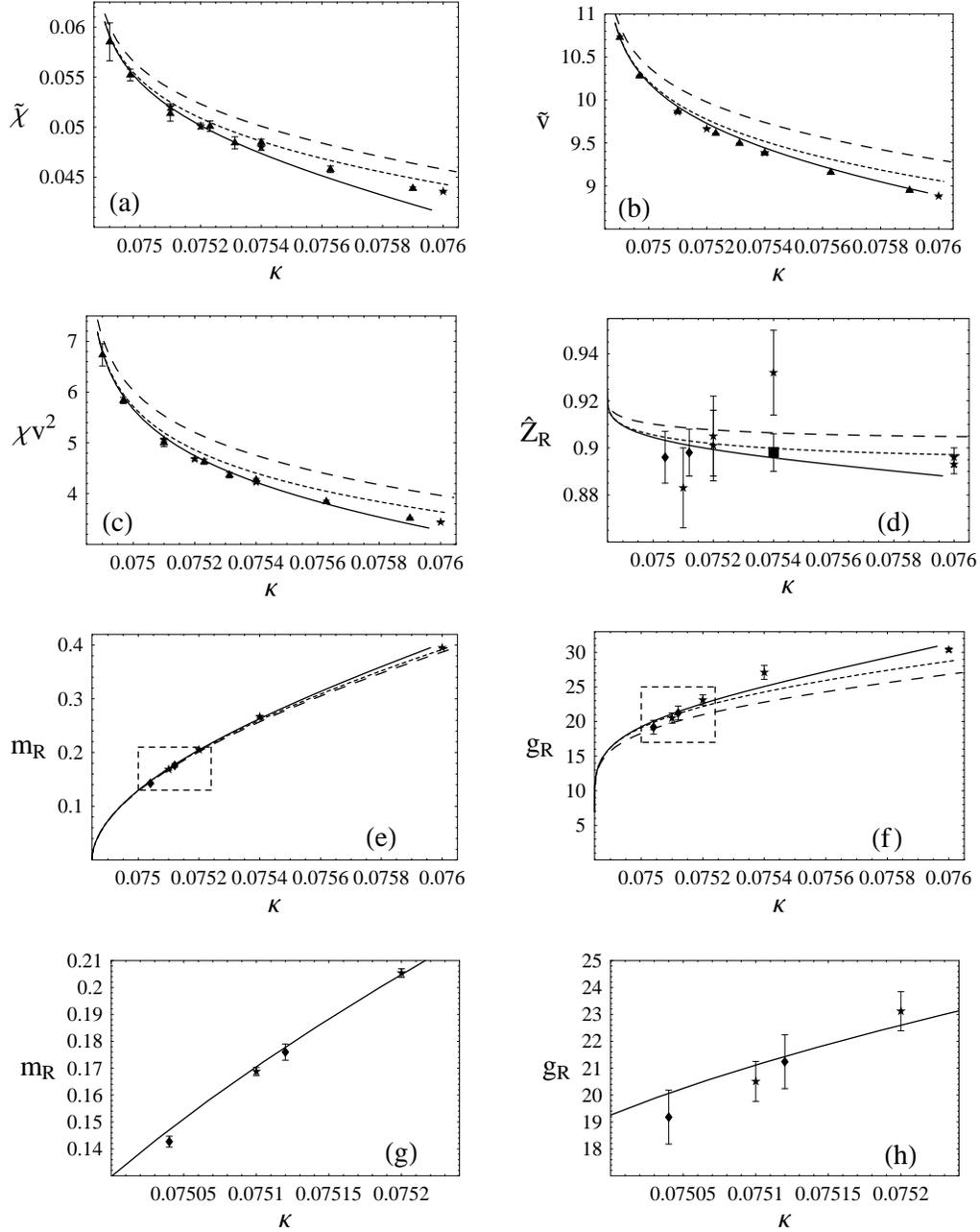}
\caption{
Broken-phase data for the Ising model compared with the LW 
procedure for parameters $\ln C_1=1.24$, $\ln C_2=1.83$, $\ln C_3=-2.90$, 
and $\kappa_c=0.074848$.  The solid curve is the full LW procedure (3-loops 
plus higher-twist corrections to 1-loop).  The long-dashed lines show the 
effect of omitting 3-loop terms, while the short-dashed lines show the effect 
of omitting higher-twist terms.  The boxed regions in (e) and (f) are shown 
on an expanded scale in (g) and (h).  Data points from Refs. \cite{Jansen,
CCC, Balog, CCCS, CCunpub}, see Appendix for details.}
\end{center}
\end{figure}  


\begin{figure}[p]
\begin{center}
\includegraphics[width=14.2cm]{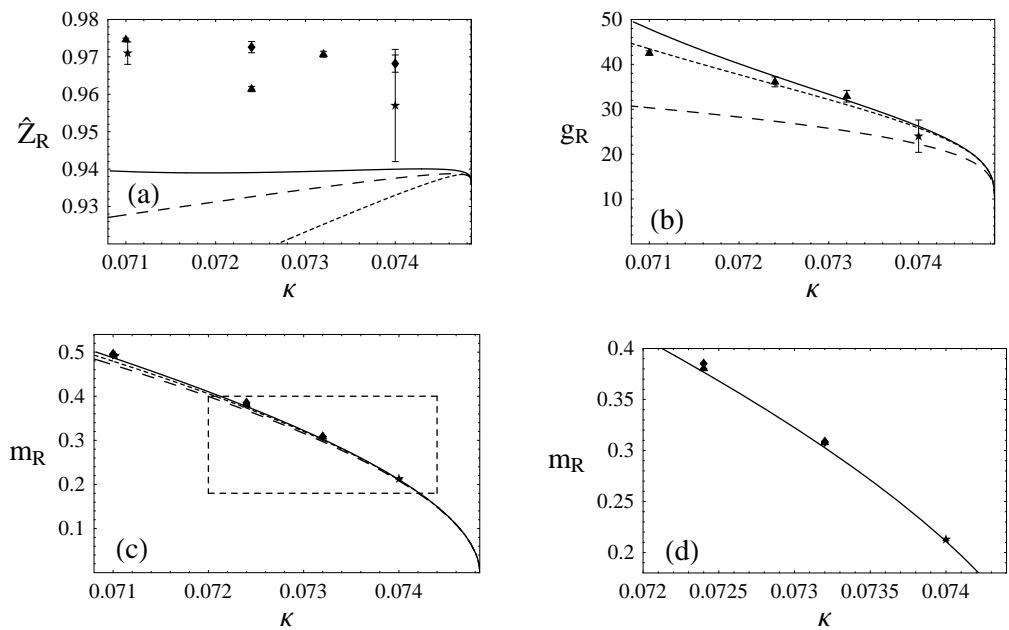}
\caption{
Symmetric-phase data for the Ising model compared with the LW 
procedure for the same parameters as Fig. 1.  Note that in (a) the curve 
lies well below the $\hat{Z}_R$ data points. The boxed region in (c) is 
shown in more detail in (d). Data points from Refs. 
\cite{MW, Montvay, CCCS, CCunpub}, see Appendix. }
\end{center}
\end{figure}  


\begin{figure}[p]
\begin{center}
\includegraphics[width=14.2cm]{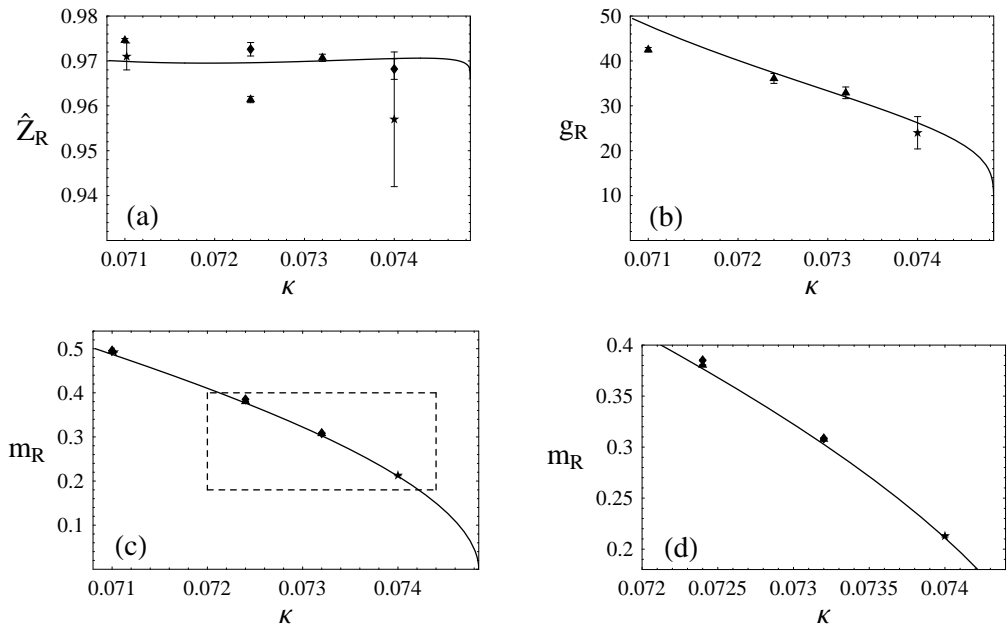}
\caption{ 
As Fig. 2 but with $\ln C_2$ increased to $1.862$ so as to obtain 
a good fit to the symmetric-phase $\hat{Z}_R$ data. }
\end{center}
\end{figure}  


\begin{figure}[p]
\begin{center}
\includegraphics[width=14.2cm]{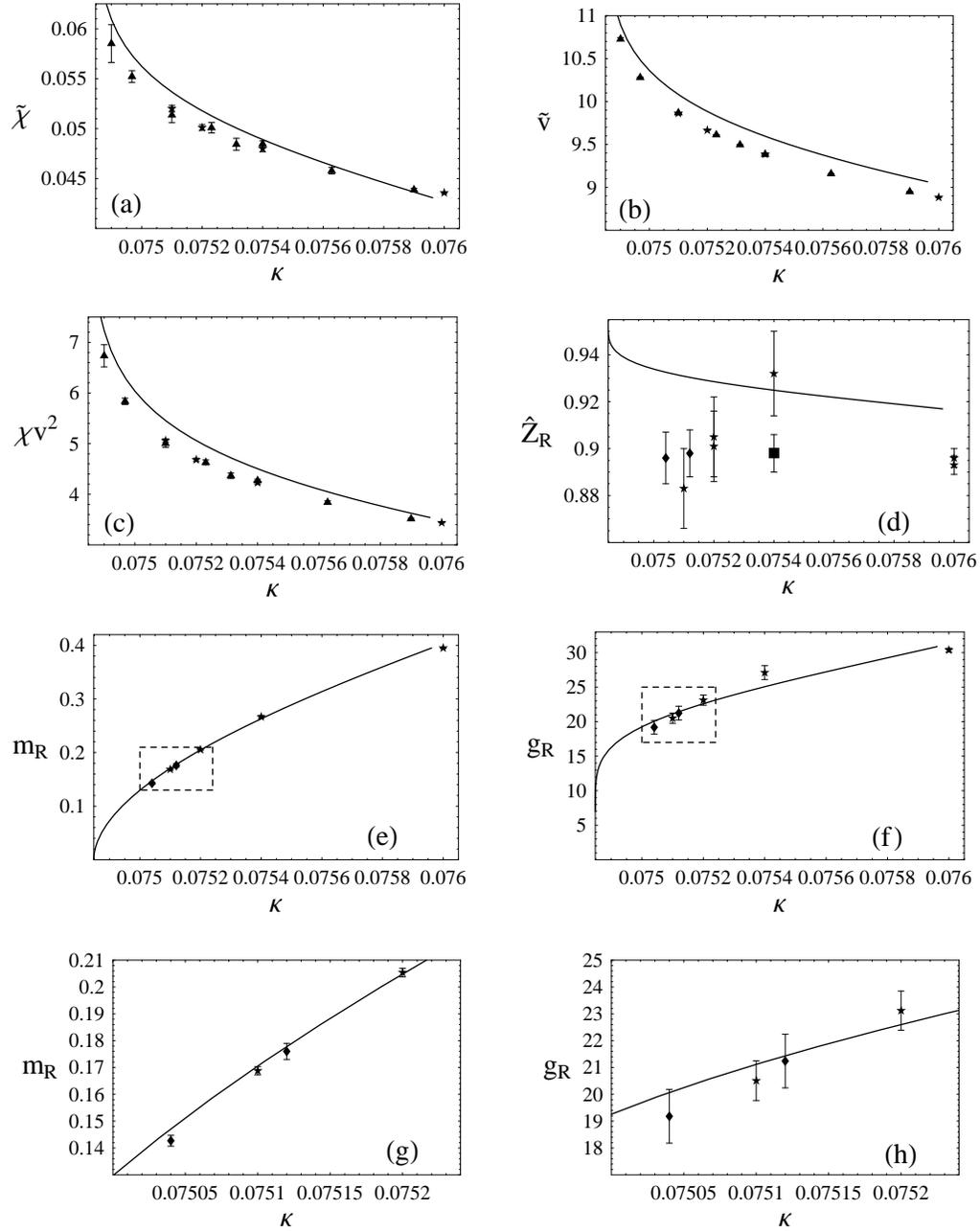}
\caption{
Fits to the broken-phase data with the same parameters as in Fig. 3: 
$\ln C_1=1.24$, $\ln C_2=1.862$, $\ln C_3=-2.90$.  Note that these parameters 
are basically in accord with LW's predicted values, Eq. (\ref{LWparam}).   
Compared to Fig. 1 the fits to $\tilde{\chi}$, $\tilde{v}$, $\chi v^2$, as 
well as $\hat{Z}_R$ in (a)--(d) are spoiled, though there is little or no 
effect on (e)--(h).}
\end{center}
\end{figure}  

\linespread{1.3}

\newpage

\section{The ``step'' in $\hat{Z}_R$}

    The essence of the problem is illustrated in Figure 5, which shows 
$\hat{Z}_R$ on {\it both} sides of the phase transition, combining Figs 2(a) 
and 1(d) on a common scale.  It is convenient to define the ``step'' as:
\BE
\Delta= \hat{Z}_R({\scriptstyle \kappa=0.074}) - 
\hat{Z}_R({\scriptstyle \kappa=0.0751}).
\EE
From the CCCS data point at $0.074$ and combining the five data points around 
$0.0751$ one finds an ``experimental'' value of $\Delta=0.071(6)$.  However, 
the theoretical curve predicts a step of only about $0.04$.  Moreover, as 
argued below, this is a robust prediction, essentially independent of the 
particular $C_i$ values, and with little theoretical uncertainty. 

\linespread{1}


\begin{figure}[htp]
\begin{center}
\includegraphics[width=11cm]{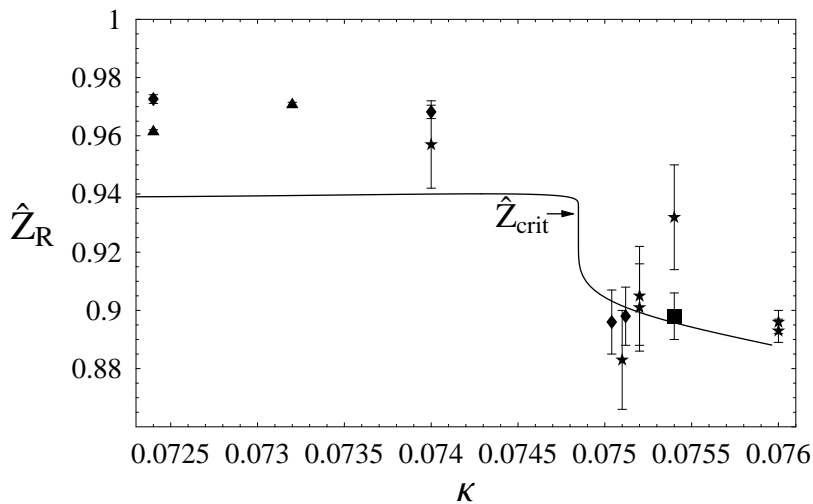}
\caption{
The ``step'' in $\hat{Z}_R$ across the phase transition.  The data 
indicate a step of $0.07$, whereas the predicted step is only $0.04$. 
The solid curve is for the parameters of Eq. (\ref{fitparam}) used in Figs. 1 
and 2.  Increasing $C_2$ and hence $\hat{Z}_{\rm crit} \equiv 2 \kappa_c C_2$ 
would shift the theoretical curve upward with almost no change in shape.  
See also Fig. 12.}
\end{center}
\end{figure}  

\linespread{1.3}

   The theoretical prediction for $\hat{Z}_R$ can be understood simply if 
we neglect higher-twist effects.  In the symmetric phase it takes the form 
\BE
\hat{Z}_R =  \hat{Z}_{\rm crit} \left(  \frac{2 \kappa}{2 \kappa_c} \right)
\left( 1+\frac{1}{18} \alpha + 0.100896 \alpha^2 + \ldots \right),
\EE
where 
\BE
\hat{Z}_{\rm crit} \equiv 2 \kappa_c C_2  \quad \quad \mbox{{\rm and}} 
\quad \,
\alpha=g_R/(16 \pi^2).
\EE
At $\kappa=0.074$, the measured $g_R$ is about $27$, so $\alpha=0.171$ and 
the series is well-behaved.  Numerically one finds, adding on a higher-twist 
contribution based on the fit in Fig. 2: 
\BE
\hat{Z}_R({\scriptstyle \kappa=0.074}) = 
\hat{Z}_ {\rm crit} (0.9887)(1.0124) + 0.0080.
\EE
[Note that in Fig. 2(a) the higher-twist contribution is quite sizeable.  
Curiously, it seems to compensate for the linear rise caused by the 
$\kappa/\kappa_c$ factor, so that $\hat{Z}_R$ is almost flat until very close 
to $\kappa_c$.]
In the broken phase one has   
\BE
\hat{Z}_R =  \hat{Z}_{\rm crit} \left(  \frac{2 \kappa}{2 \kappa_c} \right)
\left( 1-\frac{7}{36} \alpha - 0.538874 \alpha^2 + \ldots \right).
\EE
At $\kappa=0.0751$ the measured $g_R$ is about $21$, so $\alpha=0.133$ and 
again the series is well behaved.  Numerically, adding on a higher-twist 
contribution, one obtains
\BE
\hat{Z}_R({\scriptstyle \kappa=0.0751}) = 
\hat{Z}_{\rm crit} (1.0034)(0.9646) - 0.0017.
\EE
Hence, the numerical prediction for $\Delta$ is 
\BE
\Delta= 0.0331 \hat{Z}_{\rm crit} + 0.0097.
\EE
The broken-phase data suggest $\hat{Z}_{\rm crit} \approx 0.933$ while the 
symmetric-phase data imply $0.963$; in any case, $\hat{Z}_{\rm crit}$ is 
certainly less than 1.  The higher-twist contribution depends in principle 
on the $C_i$ parameters, but seems to vary little in the various fits I have 
made.  There may be some further correction from higher-loop higher-twist 
contributions, but these should be only a fraction of the higher-twist 
contribution already allowed for.  Thus, the theoretical prediction for 
$\Delta$ cannot really be pushed above $0.05$, well short of the 
``experimental'' value $0.071(6)$.    

    This is not an entirely new problem.  It was noted by Jansen {\it et al} 
\cite{Jansen} 
that there was a ``small discrepancy'' for $\hat{Z}_R$; their data point at 
$\kappa=0.076$ was about $2.5 \sigma$ below the LW prediction. (That is 
indeed what we see in  Fig. 4(d) above.)  The same problem showed up in 
Ref. \cite{VW}.  At that time it was not unreasonable to suppose that 
higher-order/higher-twist effects could explain away the discrepancy.  
However, now that there is data much closer to $\kappa_c$ that explanation 
is no longer very credible.

\section{Propagator data}

    Another indication that something unconventional may be going on comes 
from data for the momentum-space propagator.  Ref. \cite{CCCS}, 
referred to as CCCS below, found that the propagator in the broken 
phase shows significant deviations from free-propagator form.  Moreover, 
those deviations become {\it more} evident closer to $\kappa_c$, contrary 
to conventional ideas about ``triviality.''  In this section I re-examine 
that data and also point out that BDWNW's data show the same effect.   
 
Some technical preliminaries are needed. $G(\hat{p}^2)$ is defined as the 
Fourier transform of the connected two-point function.  I shall normalize 
to the {\it canonical} field $\phi_0$; hence my $G$ differs by a $2 \kappa$ 
factor from BDWNW, but agrees with CCCS.  On an $L^4$ lattice the allowed 
momenta are $p_\mu=\frac{2 \pi}{L} n_\mu$, where $n_\mu$ is a vector with 
integer-valued components and the variable $\hat{p}^2$, the lattice analogue 
of the invariant $p^{\mu}p_\mu$, is defined as $4 \sin^2(p_\mu/2)$, summed 
over $\mu=0,\ldots,4$.
  
    BDWNW do not directly provide propagator data in the Ising case 
(though they do in two non-Ising cases; see the next section).  However, 
their Table 2 gives data for the time-slice correlation function $S(t)$ 
at $\kappa=0.0751$ on a $48^4$ lattice.  From this data one can construct 
the momentum-space propagator $G(\hat{p}^2)$ for momenta 
$p=(2\pi/48)n$ ($n=0,1,2,\ldots$) along the time axis by taking the Fourier 
transform:
\[
\frac{G(\hat{p}^2)}{2 \kappa} = \sum_{t=0}^{47} S(t) e^{i p t} = 
S(0) + 2 \sum_{t=1}^{23} S(t) \cos p t + (-1)^n S(24).
\]
(The property $S(t)=S(48-t)$ has been used.  The $2 \kappa$ factor is to 
convert to the canonical normalization.)  
Without access to the raw data, I am not able to compute realistic error 
bars, but the smoothness of the data, and the good agreement with 
CCCS's data, both suggest that the error bars would be comparable to those 
of CCCS.   

 The resulting $G(\hat{p}^2)$ points are plotted in Figure 6 in various 
ways.  BDWNW's remark, at the end of Sect. 5, that ``the inverse propagator 
is remarkably linear in $\hat{k}^2$ up to the maximal (on-axis) momentum 
$\hat{k}^2=4$'' is seemingly justified by Fig. 6(b).  However, the parameters 
for this linear fit are quite different from those that BDWNW obtained from 
the low-momentum points (see Fig. 6(a)).

\linespread{1}


\begin{figure}[htp]
\begin{center}
\includegraphics[width=14.2cm]{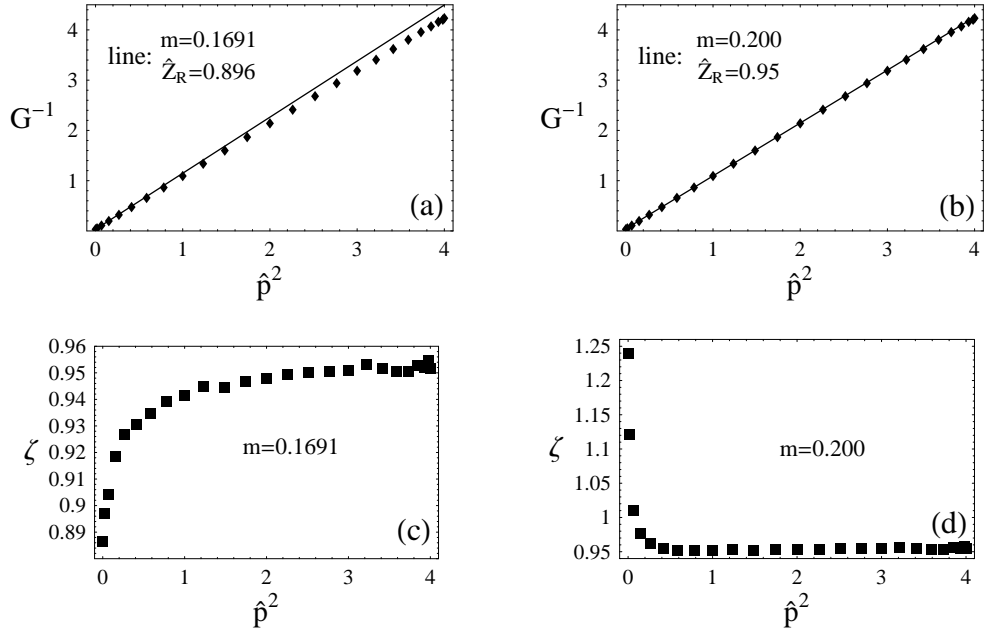}
\caption{Propagator data for $\kappa=0.0751$ from BDWNW plotted in various 
ways.  No error bars are shown. (a) shows the inverse propagator $G^{-1}$ 
with the straight line being BDWNW's free-propagator fit to the first three 
$p \neq 0$ points (see Table 5 of Ref. \cite{Balog}). (b) shows another 
straight-line fit to the same data that is good at almost all momenta.  
However, the line actually fails to fit the lowest momentum points. 
(c) and (d) show the corresponding $\zeta$ plots, 
$\zeta \equiv G(\hat{p}^2)(\hat{p}^2+m^2)$,
in which the deviations from free-propagator behaviour are more easily seen.}
\end{center}
\end{figure}  
  
\linespread{1.3}

Deviations from free-propagator behaviour are more easily seen by plotting 
the quantity
\BE
\zeta=\zeta(\hat{p}^2,m) \equiv G(\hat{p}^2)(\hat{p}^2+m^2).
\EE
Of course, $\zeta$ depends crucially on what one chooses to use as the 
``mass,'' $m$.  
Fig. 6(c) uses the $m_R=0.1691$ value given by BDWNW in Table 5, while 
Fig. 6(d) uses the larger mass, $0.200$, of the fit line in Fig. 6(b) and 
shows a $\zeta$ that is remarkably constant except for a dramatic spike at 
low momentum --- just as seen in Figs. 3, 4, 5 of CCCS \cite{CCCS}.

It is necessary to distinguish three different ``masses.''  From the 
conventional viewpoint, the physical mass is determined from the 
exponential fall-off in time of the time-slice correlator for zero 
3-momentum, $S(t)$.  I denote this mass by $m_{TS}(0)$, as in CCCS.  
The measured values of $m_{TS}(0)$ from CCCS are shown in the second column 
of Table 1.  

    In the LW procedure the ``renormalized mass'' $m_R$ and the wavefunction 
renormalization constant $\hat{Z}_R$ are defined in terms of an expansion 
of the inverse propagator about $\hat{p}^2 = 0$: 
\BE
G(\hat{p}^2)^{-1} = \hat{Z}_R^{-1} \left( m_R^2 + \hat{p}^2 + 
O(\hat{p}^4) \right).
\EE
This means that $\zeta(\hat{p}^2,m_R)$, plotted against $\hat{p}^2$, should 
have zero slope at the origin.  I have used this fact to determine $m_R$ 
empirically for the CCCS data sets at $\kappa=0.076, 0.07512, 0.07504$, 
adjusting $m$ until the lowest $\hat{p}^2$ points lined up. (See Figs. 
7, 8, 9.)  The resulting $m_R$ values, given in the third column of Table 1, 
are only slightly larger than the $m_{TS}(0)$ values.  The ratio between 
the two is quite consistent with the theoretically predicted formula 
\cite{LW2,Jansen}.  For most of the subsequent discussion the small 
difference between $m_{TS}(0)$ and $m_R$ can be ignored.  

     The third ``mass,'' denoted by $m_{\rm latt}$, corresponds, as in Figs. 
6(b,d), to the mass that gives the best fit to a free-propagator form:
\BE
G(\hat{p}^2)^{-1} \bumpeq \hat{Z}_{\rm prop}^{-1} 
\left(\hat{p}^2 + m_{\rm latt}^2 \right),
\EE
at all momenta, excepting the first few low-momentum points.  The values 
found by CCCS are given in the last column of Table 1.

\vspace*{3mm}

\begin{table}[htb]
\begin{center}
\begin{tabular}{llll}
\hline
$\kappa$ & $m_{TS}(0)$ & $m_R$ & $m_{\rm latt}$ \\
\hline
$0.076$    & $0.3912(12)$ &  $0.393$ & $0.42865(456)$ \\ 
$0.07512$  & $0.1737(24)$ &  $0.176$  & $0.20623(409)$ \\
$0.07504$  & $0.1419(17)$ &  $0.1426$ & $0.17229(336)$ \\
\hline
\end{tabular}
\end{center}

\linespread{1}

\caption{
Measured masses in the broken phase \cite{CCCS}.  The 
zero-momentum time-slice mass $m_{TS}(0)$ is the physical mass, from the 
conventional viewpoint.  The $m_R$ values, obtained as described in the 
text, differ slightly by the expected perturbative correction.  The 
$m_{\rm latt}$ values are those obtained by CCCS as giving the best fit to 
free-propagator form at all except the very lowest momenta. }
\end{table}

\linespread{1.3}

     It is important to note that in the symmetric phase the propagator 
shows no visible deviation from free-field behaviour; see Fig.~1 of CCCS 
\cite{CCCS} for $\kappa=0.074$.  All three versions of the ``mass'' are 
indistinguishable.  

    However, the situation is quite different in the broken phase.  In 
Figs. 7, 8, 9 I re-plot CCCS's propagator data using $m_R$ as the mass in 
$\zeta$.  With this mass $\zeta$ has zero slope at $\hat{p}^2=0$, by 
construction.  However, $\zeta$ then rises with $\hat{p}^2$; quickly 
at first, then more slowly.  The deviation from constancy is highly 
significant, statistically.  Moreover, the deviation from free-propagator 
behaviour is even larger at $\kappa=0.07512$ and $0.07504$, closer to 
$\kappa_c$, than it is at $\kappa=0.076$.  
 
\linespread{1}


\begin{figure}[hbtp]
\begin{center}
\includegraphics[width=10.5cm]{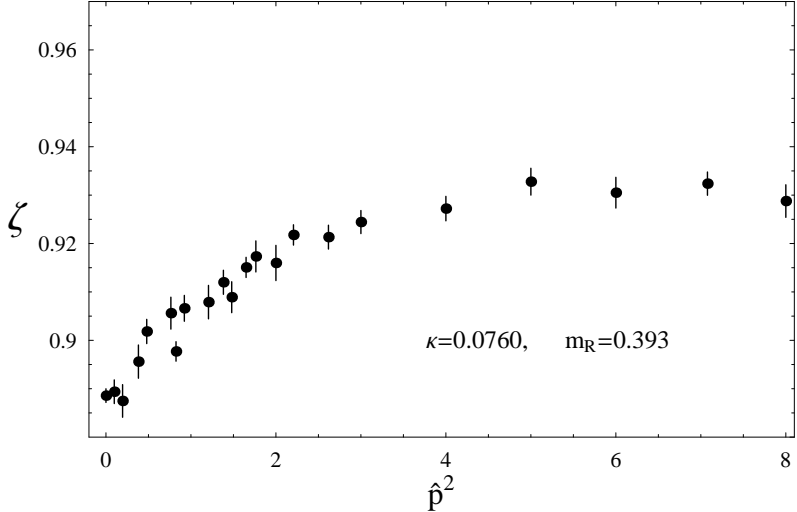}
\caption{
Propagator data for $\kappa=0.076$ from CCCS plotted as 
$\zeta \equiv G(\hat{p}^2)(\hat{p}^2+m^2)$ for a mass $m=m_R=0.393$.  
For this mass $\zeta$ has zero slope at $\hat{p}^2=0$ and its value there 
is $\hat{Z}_R$.}
\end{center}
\end{figure}


\begin{figure}[htp]
\begin{center}
\includegraphics[width=10.5cm]{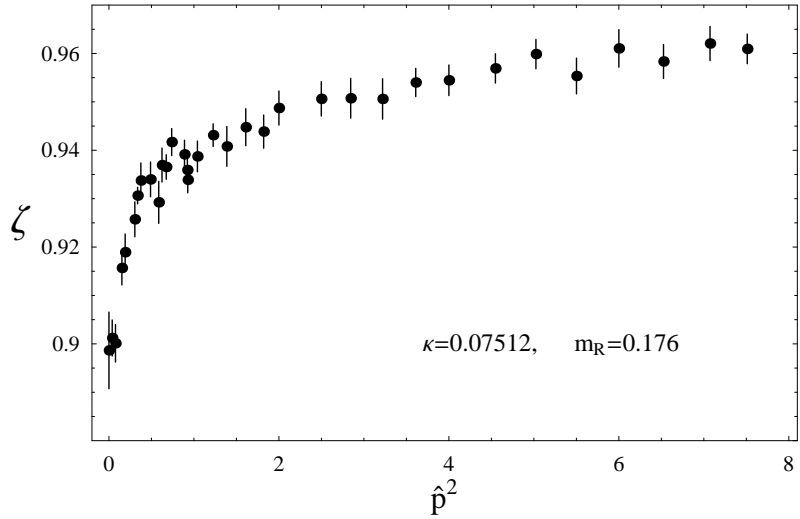}
\caption{
As Figure 7 but for $\kappa=0.07512$, with $m_R=0.176$.  The scale 
is exactly the same as Figs. 7 and 9.  Notice that 
the deviation from constancy is even greater than in Fig. 7, even though 
we are now closer to the continuum limit.}
\end{center}
\end{figure}


\begin{figure}[htp]
\begin{center}
\includegraphics[width=10.5cm]{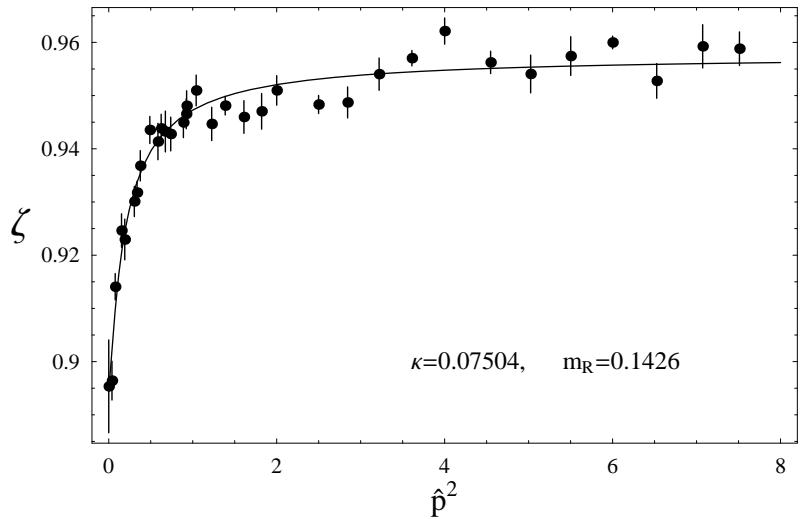}
\caption{
As Figure 7 but for $\kappa=0.07504$, with $m_R=0.1426$.  
The curve represents a simple, 
empirical fit to the data, to be used in Fig. 10.}
\end{center}
\end{figure}  

\linespread{1.3}

     While the data sets in Figs. 7, 8, 9 are exactly the same as those shown 
in Figs. 3, 4, 5 of CCCS \cite{CCCS}, the plots have a completely different 
appearance.  This fact is entirely due to the different ``m'' used in forming 
$\zeta$, just as one sees in Figs. 6(c),(d) above.  In CCCS's figures, which 
use $m_{\rm latt}$, the $\zeta$ data points are almost exactly constant, 
except for the lowest 3 or 4 points, which rise up to a dramatic peak at 
$\hat{p}^2=0$.  The peak value, $\zeta(0,m_{\rm latt})$, is CCCS's quantity 
``$Z_\phi$.''  

How this situation comes about is illustrated in Fig. 10, which uses a 
simple fit to the $\kappa=0.07504$ data.\footnote{
The fit function corresponds to 
\[
G(\hat{p}^2) = \frac{A}{\hat{p}^2+m^2_{TS}(0)} + \frac{B}{\hat{p}^2 + M^2},
\]
with parameters $A \approx 0.876$, $B\approx 0.081$ and 
$M/m_{TS}(0)$ close to 3.  This form also fits the propagator data in the 
other two cases, with quite similar values of $A,B$ and curiously, the  
best-fit $M$ is again close to $3 m_{TS}(0)$.
} 
The lowest curve corresponds to using $m=m_{TS}(0)\approx m_R$, while the 
uppermost curve corresponds to using $m=m_{\rm latt}$, chosen so that the 
curve is almost exactly flat over the whole range of $\hat{p}^2$ above, say, 
$0.1$.

\linespread{1}


\begin{figure}[htp]
\begin{center}
\includegraphics[width=14cm]{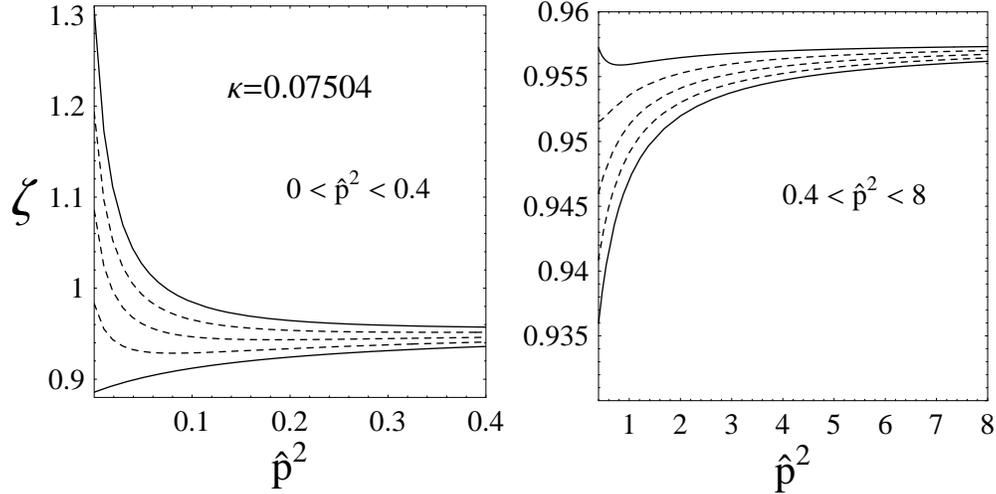}
\caption{
Illustration of the effect of plotting the propagator data as 
$\zeta \equiv G(\hat{p}^2)(\hat{p}^2+m^2)$ using different masses $m$. For 
clarity, the range of $\hat{p}^2$ is separated into $0$ to $0.4$ and $0.4$ 
to $8$.  The curves use the empirical fit to the $\kappa=0.07504$ data shown 
in the previous figure.  The lower solid curve represents $\zeta$ with 
$m=m_{TS}(0)=0.1419$, which differs only slightly from the curve 
for $m=m_R =0.1426$ shown in the previous figure.  The upper solid curve 
shows $\zeta$ with $m=m_{\rm latt}=0.17229$ used in CCCS, Fig. 5. For this 
mass $\zeta$ is essentially constant, as for a free propagator, except for 
the lowest three data points below $\hat{p}^2=0.1$.  The dashed curves 
represent intermediate choices of mass $m$ in equal steps of $m$ between 
$m_{TS}(0)=0.1419$ and $m_{\rm latt}=0.17229$.}
\end{center}
\end{figure}  

\linespread{1.3}

    CCCS argue that their data are indicative of a continuum limit in which, 
with $m_{\rm latt}$ viewed as the physical mass, the propagator tends to 
free-field form at all finite $\hat{p}^2$ except for a ``spike'' in $\zeta$ 
at infinitesimally small $\hat{p}^2$.  The top of this spike, 
$Z_\phi=\zeta(0,m_{\rm latt})$, should diverge to infinity logarithmically 
in this scenario.  Indeed, $Z_\phi$ grows from $1.05$ to $1.31$ between 
$\kappa=0.076$ and $0.07504$.  The point I want to make here is that the 
change in viewpoint as to what is the ``physical mass'' is crucial.  
Ironically, the ``odd'' features of the data --- which, from the conventional 
viewpoint are a too-low $\hat{Z}_R$ associated with a distinct {\it dip} 
in the $\zeta$ plots at low momentum --- are, from the CCCS viewpoint, 
evidence for a logarithmically growing $Z_\phi$ spike.

\section{Non-Ising data}

       BDWNW also collected some data for the $\phi^4$ theory 
not in the Ising limit, but at $\bar{\lambda}=0.3$ and $0.6$, where 
$\bar{\lambda}$ is LW's parameter that becomes unity in the Ising limit.  
The other parameters were chosen so that $g_R$ would have a value about $20$, 
for comparison with their Ising data at $\kappa=0.0751$.  Their propagator 
data at $\bar{\lambda}=0.3, 0.6$ show no statistically significant deviation 
from constancy, even if re-plotted using $\zeta$.  (However, their data only 
extend to $\hat{p}^2=0.3$ and possibly some effect might show up if one 
had data at higher momenta.)

\linespread{1}


\begin{figure}[htp]
\begin{center}
\includegraphics[width=10cm]{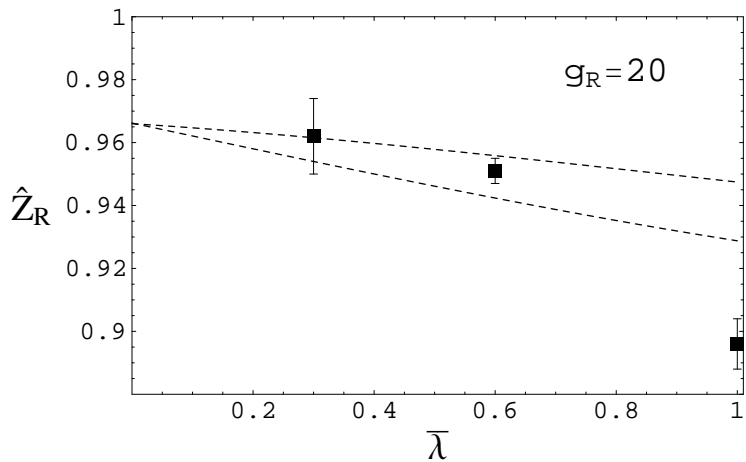}
\caption{
$\hat{Z}_R$ in $\phi^4$ theory at $g_R =20$, as a function of the 
parameter $\bar{\lambda}$.  The region between the two dotted curves 
represents the prediction of LW(II).  The three bold points are the data 
points from BDWNW, Table 5.  A deviation is seen only in the 
Ising case, $\bar{\lambda}=1$. }
\end{center}
\end{figure}  

\linespread{1.3}

Figure 11 plots the BDWNW results for $\hat{Z}_R$ in comparison with 
the LW expectation, indicated by the region between the two dotted curves.  
These curves were obtained by finding 
$\hat{Z}_{\rm crit} \equiv 2 \kappa_c C_2$ from the $C_2$'s of LW(II), 
Table 1, and the $\kappa_c$'s of LW(I), Table 1, and then applying the 
perturbative correction, $0.967$, relating $\hat{Z}_R$ to 
$\hat{Z}_{\rm crit}$ at $g_R=20$.  
The moral of this plot is that the $\hat{Z}_R$ problem discussed earlier 
appears to show up only in the Ising case; i.e., it becomes visible only for 
$\bar{\lambda}$ above $0.6$.

\section{Summary and Conclusions}

      In many respects the RG predictions of LW are impressively successful; 
they explain a large amount of data over a sizeable range.  However, on 
close examination, there does appear to be a significant problem:
The parameters that fit the broken-phase data well (Fig. 1) do not fit the 
symmetric-phase data for $\hat{Z}_R$ (Fig. 2(a)).  Alternatively, parameters 
that fit the symmetric-phase data well (Fig. 3), and which accord well with 
LW's predicted values, do not fit the broken-phase data (Fig. 4).  The core 
of the problem is that the data require a downward step in $\hat{Z}_R$ of 
$0.07$ across the phase transition, whereas the theory predicts a step 
of only about $0.04$.  This is a serious concern because it is the proudest 
boast of the RG method that it can relate the behaviours on each side of the 
phase transition.  

\linespread{1}


\begin{figure}[htp]
\begin{center}
\includegraphics[width=10cm]{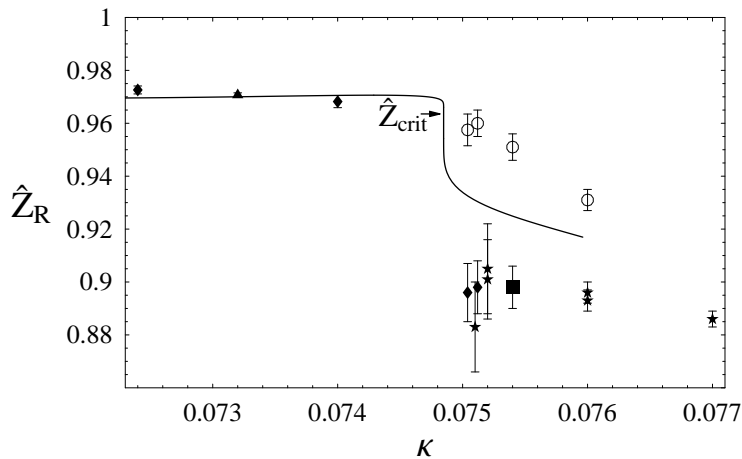}
\caption{The ``step'' in $\hat{Z}_R$ across the phase transition, revisited.  
The solid curve is the RG prediction using the parameters of Figs. 3 and 4 
that fit the symmetric-phase data well.  The open circles represent 
``$\hat{Z}_{\rm prop}$'' determined from the $\zeta$ values of the 
large-$\hat{p}^2$ propagator data.  I have used some ``editorial license'' to 
omit some data points that I consider dubious or uninformative (see discussion 
in the Appendix).}
\end{center}
\end{figure}  

\linespread{1.3}

    Figure 12 represents my best summary of the situation, expressed in 
conventional terms.  The RG curve here corresponds to $C_i$ parameters chosen 
to give a good fit to the {\it symmetric}-phase data.  Relative to this RG 
prediction, the measured $\hat{Z}_R$'s in the broken phase are too low.  
The open circles correspond to ``$Z_{\rm prop}$'' determined from the 
large-$\hat{p}^2$ behaviour of the propagator data.  These points appear to 
join smoothly to the symmetric-phase points (where $\hat{Z}_R$ and 
$\hat{Z}_{\rm prop}$ are indistinguishable).  In the broken phase, the 
vertical gap between the open circles and the $\hat{Z}_R$ data points is a 
measure of the deviation from free-propagator behaviour.  Note that this gap 
{\it increases} as one approaches $\kappa_c$.  

    Because of the effect illustrated in Fig. 10 these features of the data 
can be re-interpreted, from CCCS's viewpoint, in terms of a logarithmically 
growing $Z_\phi$.  From the conventional viewpoint, one can only say 
that there is a puzzle that remains to be resolved.  My conclusion is that 
there is strong motivation for collecting more Ising-model lattice data, 
especially closer to $\kappa_c$, on both sides, where the residual 
uncertainties from higher-twist effects will be even smaller. 

\vspace*{15mm}

\hspace{-\parindent}
{\bf Acknowledgements}

\vspace*{3mm}

I most grateful to Paolo Cea and Leonardo Cosmai for permission to quote 
some of their new, unpublished data.  I thank them and Maurizio Consoli 
for many discussions.  
I am also grateful to Anthony Duncan and Ray Willey for helpful discussions 
about the BDWNW results.  
This work was supported in part by the Department of Energy under Grant No. 
DE-FG05-97ER41031.

\newpage

\section*{Appendix: Data}

    The broken-phase data in Fig. 1 come from various sources.  Triangles 
($\blacktriangle$) represent data for $\chi$ and $v$ from CCC \cite{CCC}.  
Stars ($\bigstar$) represent data from BDWNW \cite{Balog} and also an 
earlier data point from Jansen {\it et al} \cite{Jansen} at $\kappa=0.076$.  
A convenient compilation of these data points can be found in Table 3 of 
BDWNW \cite{Balog}.  Diamonds ($\blacklozenge$) represent data at 
$\kappa=0.07512$ and $0.07504$ for $\hat{Z}_R$, $g_R$, and $m_R$ that I 
have extracted from results of CCCS Ref. \cite{CCCS}.  (The CCCS results at 
$0.076$ agree completely with Jansen {\it et al}.)   

    The symmetric-phase data in Fig. 2 also come from various sources.  
Triangles ($\blacktriangle$) represent data from Montvay, M\"unster, and 
Wolff \cite{Montvay}.  Stars ($\bigstar$) represent earlier data from 
Montvay and Weisz \cite{MW}.  The diamond ($\blacklozenge$) at $\kappa=0.074$ 
comes from CCCS \cite{CCCS}.  The diamond at $\kappa=0.0724$ represents 
unpublished data of Cea and Cosmai \cite{CCunpub}; see comments below.

    In general there appears to be very satisfactory agreement between the 
data of the various groups.  (To avoid clutter I have generally not 
plotted a data point that completely agrees with, but is less precise than, 
the equivalent data point from another group.)  A few data points, however, 
deserve comment because superficially they might appear to weaken my case 
for a large step in $\hat{Z}_R$.  

    The BDWNW data point for $\hat{Z}_R$ at $\kappa=0.0754$ appears high in 
comparison with the others.  I believe that the most likely explanation 
is an unlucky $2 \sigma$ statistical fluctuation.  I say this (and it implies 
no criticism of BDWNW) for two reasons: (i) $\hat{Z}_R$ in the region 
$0.075<\kappa<0.076$ is expected to vary slowly and smoothly, so at least 
one of the BDWNW points must be shifted by more than $1.5 \sigma$.  Since 
the other BDWNW points are well corroborated by independent data, the 
$0.0754$ point is the likely culprit.  (ii) The $0.0754$ data point also 
seems to lie slightly off the fit curves in the $m_R$ and $g_R$ plots 
((e) and (f) in either Fig. 1 or Fig. 4); the hypothesis that the $m_R$ value 
has a $2 \sigma$ upward fluctuation would explain away all the discrepancies. 

     In fact, after having written the previous paragraph, I learned of 
new data of Cea and Cosmai \cite{CCunpub} at the same $\kappa$, which indeed 
yield a smaller mass, $m_R=0.262(1)$ and a lower $\hat{Z}_R=0.898(8)$.  This 
result is included in the $\hat{Z}_R$ figures as a square ($\blacksquare$). 
(To avoid clutter it has not been included in the $m_R$ plots.)   

     In the symmetric phase, the $\kappa=0.0724$ result of Montvay 
{\it et al} \cite{Montvay} for $\hat{Z}_R$ appears to be considerably lower 
than its two neighbouring points, also from Montvay {\it et al}.  The quoted 
errors are very small, so statistics should not be a factor.  The predicted 
$\hat{Z}_R$ in this region is almost exactly constant, so the sharp dip 
implied by the three Montvay {\it et al} ($\blacktriangle$) points 
(see Fig 2(a) or 3(a)), 
if it were real, would be in serious disagreement with theory.  Because of 
this anomaly I asked Cea and Cosmai to repeat the Monte-Carlo calculation 
of $\hat{Z}_R$ at the three $\kappa$ values studied by Montvay {\it et al}.  
They kindly did so \cite{CCunpub} and found excellent agreement at $0.0710$ 
and $0.0732$, but at $0.0724$ they found a different result that indeed 
interpolates smoothly between the neighbouring $\kappa$'s.  I therefore 
suspect that there must be some trivial mistake in Montvay {\it et al}'s 
$0.0724$ point -- perhaps in the transcription of the actual computer result 
to the published paper.  

    The symmetric-phase data point at $\kappa=0.074$, closest to the phase 
transition, is obviously important.  The Montvay-Weisz (MW) data point 
($\bigstar$) for $m_R$ comes from the last line of Table 2a of \cite{MW} as 
$0.2125(10)$ times a factor $1.002$ (see last sentence of Sect. 3.1) to 
convert the physical (or ``time-slice'') mass to $m_R$.  This gives 
$0.2129(10)$, which agrees very well with CCCS's result $0.2141(28)$ (not 
shown in the figures).  The MW result for $\hat{Z}_R$ also comes from the 
last line of Table 2a of MW, from ``$z=6.44(10)$'' converted by a 
$2 \kappa$ factor and $(1.002)^2$ (see Eqs. (26), (27) in MW), giving 
$\hat{Z}_R=0.957(15)$.  This is compatible with the more precise value from 
CCCS, $0.9682(23)$ obtained from propagator data.  The $g_R$ value from 
MW comes from remarks in their Sect. 4.2 that a $g_R$ of $24$ with a 
$10-15\%$ possible variation would fit their data.

\end{document}